\documentstyle[prl,aps,epsf]{revtex}
\begin{document}

\advance\textheight by 0.5in
\advance\topmargin by -0.25in
\draft

\twocolumn[\hsize\textwidth\columnwidth\hsize\csname@twocolumnfalse%
\endcsname

\preprint{NSF--ITP--97--104, cond-mat/9708054}       

%\tighten
\title{\hfill {\small ITP Preprint Number NSF--ITP--97--104,
    cond-mat/9708054}  \\ 
\vspace{10pt}
Coulomb Interactions and Mesoscopic Effects in Carbon Nanotubes}
\author{Charles Kane}
\address{Department of Physics, University of Pennsylvania\\
Philadelphia, Pennsylvania 19104
}

\author{Leon Balents and  Matthew P. A. Fisher}
\address{Institute for Theoretical Physics, University of California, 
Santa Barbara, CA 93106--4030}

\date{\today}
\maketitle

%\widetext

\begin{abstract}
  We argue that long-range Coulomb forces convert an isolated ($N,N$)
  armchair carbon nanotube into a strongly-renormalized {\sl Luttinger
    liquid}.  At high temperatures, we find anomalous temperature
  dependences for the interaction and impurity contributions to the
  resistivity, and similar power-law dependences for the local
  tunneling density of states.  At low temperatures, the nanotube
  exhibits spin-charge separation, visible as an extra energy scale in
  the discrete tunneling density of states (for which we give an
  analytic form), signaling a departure from the orthodox theory of
  Coulomb blockade.  
\end{abstract}
\vskip -0.5 truein
]
%\twocolumn
%\narrowtext

The rapid experimental progress in controlled preparation of long,
single-walled nanotubes bodes well both for applications and
fundamental science\cite{Review}.  Recent proposals for their use
include tips for scanning microscopy, ultra-strong mechanical fibers,
pinning sites for high-T$_c$ superconductors, and inclusions in
composites for body armor!  One of the most exciting prospects from
the point of view of physics is that of a nearly ideal quantum wire.
``Buckytubes'' promise to be smaller, longer, cleaner and more
chemically manipulable than their semiconducting or metallic
counterparts.  For this purpose, probably isolated single-walled
nanotubes are most relevant, and a thorough understanding of their
electronic properties is desirable.

Previous papers by various authors have discussed the band-structures
in various geometries\cite{Theory1,Theory2}, as well as the effects of
the electron-phonon\cite{Twiston}\ and short-range (Hubbard-like)
electron-electron interactions\cite{Balents97,Krotov97}.  These types of
interactions have only weak effects, leading to a small linear
correction to the resistivity at high temperatures\cite{ropes}, more
significant deviations from non-interacting behavior occurring only at
a very low temperature scale of order $E_F e^{-N}$, where $E_F$ is the
Fermi energy and $N$ is the circumference of the tube in units of the
graphene periodicity.  While these treatments may be appropriate for
arrays (``ropes''), they are inadequate for isolated nanotubes, due to
the unscreened nature of the Coulomb interaction in this situation.

In this letter, we address the effects of the long-range Coulomb
potential on the most conducting ``armchair'' tubes.  Once
these are included, we find that significant deviations from
non-interacting behavior should be observable {\sl at all
  temperatures}.  At high temperatures, an isolated armchair nanotube
should behave as a Luttinger liquid, with an anomalous power-law
dependence of the resistivity and power-law tunneling density of
states, scaling differently at the end and center of the tube.  At low
temperatures, Coulomb blockade behavior sets it, but with considerable
deviations from the ``orthodox theory''\cite{TunnelingExpts}.  In
particular, the conductance peak spacing is characterized by {\sl
  three} energy scales in contrast to the usual two.  In addition to
the usual charging energy $E_C$ and single-particle level spacing
$\epsilon_0$, a third energy scale $\varepsilon_\rho$ reflects the separation of
spin and charge in the 1d Luttinger liquid.  Furthermore, a
non-trivial ratio of peak heights is expected, arising from the
collective nature of the low-energy excitations and invalidity of the
quasiparticle picture.  A full quantitative expression (for the
tunneling density of states of a finite length nanotube) containing
this physics, which holds away from half-filling and for sufficiently
short tubes in the undoped case, is given in Eq. 10.  A discussion of
the experimental situation is given at the end the letter.

%The remainder of the paper is organized as follows.  We first review
%the basic band-theory of the conducting ``armchair'' nanotubes,
%describing the consequent Luttinger model and its bosonized form.
%Next, we introduce the Coulomb potential, and show that it leads to
%small ($O(1/N)$) backscattering and Umklapp interaction, but a {\sl
%  large} forward scattering amplitude.  This forward scattering term
%is then treated {\sl exactly} using bosonization, leading to the
%Luttinger liquid model anticipated above.  Using this model, we next
%consider two regimes.  At high temperatures, the finite length of the
%nanotube can be neglected, and we derive for this regime the
%resistivity and continuous density of states.  At lower temperatures
%for infinite tubes, the weak Umklapp and backscattering terms will
%ultimately drive an instability of the Luttinger liquid, which is
%beyond the scope of this letter.  This instability is however cut off
%by the discrete level spacing in a finite length nanotube, and we
%next discuss  the temperature and length scales for which the
%Luttinger description continues to hold.  Assuming these are
%satisfied, we then derive the appropriate {\sl discrete} density
%of states for the finite Luttinger liquid, which governs the
%conductance of the nanotube in the Coulomb blockade regime.

We begin by reviewing the band structure of the (N,N) armchair tube,
which has been discussed by several authors.  It is well-captured by a
simple tight-binding model of $p_z$ electrons on the honeycomb
lattice.  For the armchair tubes, evaluating the resulting
tight-binding band structure for the discrete set of allowed quantized
transverse momenta $q_y$ leads to only two gapless {\sl
  one-dimensional} metallic bands (with $q_y =
0$).\cite{Theory1,Theory2}\ These 
dominate the low-energy physics, disperse with the same velocity,
$v_F$, and can be described by the simple 1d free $ $Fermion model,
\begin{equation}
H_0 = \sum_{i,\alpha} \int dx v_F \left[
    \psi^{\dag}_{Ri\alpha}i\partial_x \psi^{\vphantom\dag}_{Ri\alpha}
    -\psi^{\dag}_{Li\alpha}i\partial_x \psi^{\vphantom\dag}_{Li\alpha}
  \right],  
\label{H_0}
\end{equation}
where $i=1,2$ labels the two bands, and $\alpha = \uparrow,\downarrow$
the electron spin.  

We will make heavy use of the bosonized representation of
Eq.~\ref{H_0}, obtained by writing $\psi_{R/L;i\alpha} \sim
e^{i(\phi_{i\alpha} \pm \theta_{i\alpha})}$, where the dual fields
satisfy $[\phi_{i\alpha}(x), \theta_{j\beta}(y)] =
-i\pi \delta_{ij}\delta_{\alpha\beta} \Theta(x-y)$.  
Expressed in these variables (1) takes the form
$H_0 = \sum_{i,\alpha} {\cal H}_0(\theta_{i\alpha},\phi_{i\alpha})$
\begin{equation}
{\cal H}_0(\theta,\phi) = \int dx \, {v_F \over {2\pi}} [(\partial_x
\theta)^2 + (\partial_x \phi)^2].
\label{Hspin}
\end{equation}
The slowly varying electronic density in a given channel is given by
$\rho_{i\alpha} \equiv \psi_{Ri\alpha}^\dagger
\psi_{Ri\alpha}^{\vphantom\dag} + \psi_{Li\alpha}^\dagger
\psi_{Li\alpha}^{\vphantom\dagger} = \partial_x\theta_{i\alpha}/\pi$.
The normal modes of ${\cal H}_0$ describe long wavelength particle-hole 
excitations which propagate with a dispersion $\omega = v_F q$.

%Bosonic Euclidean action is then simply,
%\begin{equation}
%  S_0 = \sum_{i\alpha} \int dx d\tau {\cal
%    L}_0(\theta_{i\alpha},\phi_{i\alpha}) , 
%\end{equation}
%with
%\begin{equation}
%  {\cal L}_0(\theta,\phi) = {v_F \over {2\pi}} [(\partial_x \theta)^2 +
%  (\partial_x \phi)^2] + {i \over \pi} \partial_x \phi \partial_\tau \theta .
%\end{equation}
%Upon integrating out the $\phi$ field, this can equivalently be
%written,
%\begin{equation}
%  {\cal L}_0(\theta) = {1 \over {2\pi}} [v_F (\partial_x \theta)^2 +
%  {1 \over v_F} (\partial_\tau \theta)^2 ] .
%\end{equation}
%The slowly varying electron density, $\rho = \psi^\dagger_R \psi_R +
%\psi^\dagger_L \psi_L$, has the simple form $\rho = \partial_x
%\theta/\pi$ when bosonized, so this describes a density wave
%propagating at the Fermi velocity.

Turning to the interactions, a tremendous simplification occurs when
$N$ is large: the only couplings which survive in this limit are
{\it forward scattering} processes which involve small momentum
transfer.  Roughly speaking, this can be understood as follows.
``Interbranch" scattering processes (such as backscattering and
umklapp) involve a momentum transfer of order $2k_F \sim 1/a$, where
$a$ is the carbon-carbon bond length.  The matrix elements are
therefore dominated by the {\it short range} part of the interaction,
at distances $r\sim a$, where the interaction changes significantly
from site to site.  However, the electrons in the lowest sub-band are
spread out around the circumference of the tube, and for large $N$ the
probability of two electrons to be near each other is of order $1/N$.
For the Coulomb interaction, the resulting dimensionless interaction
vertices are of order $(e^2/h v_F) \times 1/N$\cite{Balents97,Twiston}.  By
contrast forward scattering processes, in which electrons stay in the
same branch, involve small momentum exchange.  They are dominated by the
{\it long range} part of the Coulomb interaction, at distances larger
than the radius, and there is no $1/N$ suppression.

For $N \gtrsim 10$ it is thus appropriate to consider a {\it Luttinger
  model}, in which only forward scattering vertices are included.  A
further simplification arises because the {\it squared} moduli of the
electron wavefunctions in the two bands are {\it identical} and spin
independent.  All the forward-scattering vertices can thus be written
as a {\it single} interaction, coupling to the total charge density
$\rho_{\rm tot} = \sum_{i\alpha} \partial_x \theta_{i\alpha}/\pi$.

We will suppose that the Coulomb interaction is externally screened on 
a scale $R_s$, which is long compared to the tube radius $R$, but
short compared to the length of the tube.
For simplicity, we model this by a metallic cylinder of
radius $R_s$, placed around the nanotube.  From elementary
electrostatics, the energy to charge the
nanotube with an electron density $e\rho_{tot}$ is
\begin{equation}
  H_{\rm int} = e^2 \ln(R_s/R) \int dx  \rho_{\rm tot}^2  .
\end{equation}

Since $H_{\rm int}$ only involves $\rho_{\rm tot}$ it is
convenient to introduce a spin
and channel decomposition via, $\theta_{i,\rho/\sigma} =
(\theta_{i\uparrow} \pm \theta_{i\downarrow})/\sqrt{2} $ and
$\theta_{\mu \pm} = (\theta_{1\mu} \pm \theta_{2\mu})/\sqrt{2}$ with
$\mu = \rho, \sigma$, and similar definitions for $\phi$.  As defined,
the new fields $\theta_a$ and $\phi_a$ with $a=(\rho/\sigma, \pm)$,
satisfy the same canonical commutators $[\phi_a(x), \theta_b(y)] =
-i\pi\delta_{ab} \Theta(x-y)$.  
In the absence of interactions the Hamiltonian is simply
$H_0 = \sum_a \int_{x,\tau} {\cal H}_0(\theta_a,\phi_a)$. 
which describes three ``sectors" of neutral excitations and one
charged excitation.  Including the interactions only modifies
the charge sector, which is described by the sum of two terms
${\cal H}_\rho = {\cal H}_0(\theta_{\rho+},\phi_{\rho+}) 
+ H_{\rm int}(\theta_{\rho+})$, and may be written
\begin{equation}
{\cal H}_\rho = \int dx {v_\rho\over {2\pi }} \left[
  g^{-1}(\partial_x\theta_{\rho+})^2 
+ g  (\partial_x\phi_{\rho+})^2 \right].
\label{Hcharge}
\end{equation}
This describes the 1d acoustic {\it plasmon} which propagates with velocity
$v_\rho = \sqrt{v_F(v_F + (8e^2/\pi \hbar) \ln (R_s/R))}$ and is characterized
by the Luttinger parameter $g = v_F/v_\rho$.

%The latter is in fact just the 1d plasmon, since
%the total density is $\rho_{\rm tot}= (2/\pi) \partial_x
%\theta_{\rho+}$.

%Expressed in these new variables, we have 
%$H_0 + H_{\rm int} = \sum_a \int dx {\cal H}_a$ with
%\begin{equation}
%{\cal H}_a = {v_a\over 2\pi} \left[ {1\over g_a} (\partial_x\theta_a)^2
% + g_a (\partial_x\phi_a)^2 \right]
%\end{equation}

%Upon bosonization, this term only
%modifies the plasmon mode, which is described by a sum of two terms,
%$S_{\rho} = \int_{x,\tau} {\cal L}_0(\theta_{\rho+}) +
%S_{int}(\theta_{\rho+})$, and can be written,
%\begin{equation}
%  S_{\rho} = {1 \over {2\pi g}} \int_{x,\tau} [v_\rho (\partial_x
%  \theta_\rho)^2 + {1 \over v_\rho} (\partial_\tau \theta_\rho)^2 ]  ,
%\end{equation}
%with an increased plasmon velocity, $v_\rho = v_F/g$, and a modified
%``Luttinger" stiffness parameter, $g = \sqrt{v_F/(v_F + v_c)}$, with a
%``Coulomb" velocity defined as $v_c = (8e^2/\pi \hbar) \ln(R_s/R)$.

With repulsive forward scattering interactions, the plasmon velocity
$v_\rho$ is larger than $v_F$, exhibiting the well known spin-charge 
separation of a 1d Luttinger liquid.  Moreover, the Luttinger 
parameter, $g$, which
equals one in a Fermi liquid is reduced.  Since these effects are
coming from long-ranged Coulomb forces and the short-ranged
contributions are smaller - down by $1/N$ - it is possible to make
{\it semi-quantitative} estimates of Luttinger liquid effects.
Specifically, with a Fermi velocity estimated from graphite
bandstructure of $v_F = 8 \times 10^5$ m/s and a screening length
of, say, $1000$\AA\  one finds $g \approx 0.2$ -
well below the Fermi liquid value, $g=1$.  This result is relatively
insensitive to the screening, depending only logarithmically on the
length $R_s$.

Physical properties can be readily evaluated from
Eqs.~\ref{Hspin},\ref{Hcharge}.  For example, consider the density of
states to tunnel an electron into a long nanotube from a metallic
electrode or perhaps an STM tip.  Upon expressing the electron
operator in terms of the boson fields and evaluating the electron
Green's function one finds $\rho_{\rm tun}(\epsilon) \sim
\epsilon^{\alpha}$, with an exponent $\alpha = (g + g^{-1} -2)/8$,
which vanishes in the Fermi liquid limit ($g=1$), but is expected to
be quite appreciable, $\alpha \approx 0.4$ for the nanotube.  The
resulting tunneling current should be suppressed with, $dI/dV \sim
V^\alpha$, and the linear conductance $G(T) \sim T^\alpha$ vanishing
with temperature.  This suppression is even more dramatic for
tunneling into the {\it end} of a long nanotube.  One finds a larger
exponent, $\alpha_{\rm end} = (g^{-1} -1)/4 \approx 1$.

These results were established under the assumption that the backward
and umklapp interactions could be safely ignored.  Since their bare
values are small, of order $1/N$, this might seem very reasonable.
However, the presence of the strong forward scattering greatly
modifies the effects of the Umklapp scattering at low energies, so
caution is necessary.  To estimate this effect we reconsider the
neglected interactions as perturbations upon the Luttinger model. We
find that the momentum-conserving backward scattering vertices,
$u_{bs}$, are ``marginal".  They only become important at an
exponentially small energy scale, $\Delta_{bs} \sim E_F
\exp(-c/u_{bs})$ with an order one constant, $c$.  At half-filling,
however, the umklapp scattering vertices $u$ grow much more rapidly at
low energies due to the stiff plasmon mode.  Their renormalized
strengths at energy $\epsilon$ grow as $u(\epsilon) \sim u
(E_F/\epsilon)^{1-g}$.  This growth signifies the development of a
gap in the spectrum, with magnitude $\Delta \sim E_F u^{1/(1-g)}$.  The
above Luttinger liquid results are only valid on energy scales well
above this gap, where the Umklapp scattering can still be safely
ignored.  Unfortunately, a reliable quantitative estimate for this gap
is difficult.  For nanotubes doped away from $1/2$ filling, umklapp
processes suffer a momentum mismatch at the $ $Fermi surface, thereby
becoming ineffective.  In particular, for a doped tube with Fermi
energy shifted away from the band center by an energy $\delta$ which
satisfies $u(\delta) \ll 1$ ($\delta \gg E_F/N^{1/(1-g)}$ for
$e^2/\hbar v_F$ of order one), the validity of the Luttinger model
should be limited only by the exponentially low backscattering scale
$\Delta_{bs}$.

In an undoped tube, at temperatures above the energy gap, one expects
the umklapp interactions to cause weak backscattering and lead to a
non-zero resistivity.  The resistivity should be proportional to the
electron backscattering rate, varying as $\rho(T) \sim u(T)^2 T$ where
$u(T) \sim T^{g-1}$ is the energy (temperature) dependent umklapp
scattering strength.  The resulting non-linear power-law behavior,
$\rho(T) \sim T^{2g-1}$, valid over a temperature range above the gap,
is a clear signature of the Luttinger liquid.  For temperatures below
the gap, this should crossover into an activated form, diverging
exponentially as $T \rightarrow 0$.  Backscattering mediated by
twiston phonons (if unpinned by the substrate) should lead to the same
temperature dependence.  Impurities also have dramatically
enhanced effects in the Luttinger liquid.  Like umklapp and twiston
scattering, disorder leads to a high-temperature power-law resistivity, but
with $\rho(T) \sim T^{-(1-g)/2}$.

The above discussion has implicitly assumed that the nanotube is {\it
  infinitely} long.  For a tube with finite length, $L$, many
interesting mesoscopic effects are expected.  For temperatures and/or
voltages well above the level spacing $\pi \hbar v_F/L$, the above results
(for $L = \infty$) should be valid.  We now turn to a discussion of
the mesoscopic effects on smaller energy scales.
For simplicity, we assume that $L$ is sufficiently small so that the
energy gaps induced by umklapp and backward scattering satisfy
$\Delta,\Delta_{bs} \ll \hbar v_F/L$.  In this limit, it is valid to
employ the Luttinger model.

For a finite tube it is
convenient to express $\theta_a,\phi_a$
in terms of creation and annihilation operators for the discrete
bosonic excitations.  At the tube ends, these fields must satisfy
the boundary conditions $\partial_x \phi(x=0,L) = 0$ and
$\theta_a(L) - \theta_a(0) = (\pi N_a + \delta_a)/2$,
where $N_a$ is the integer charge in the $a$ sector and
$\delta_a$ is a sum of phase shifts associated with the tube
ends\cite{phaseshift}. 
Expanding in a Fourier series gives,
\begin{equation}
\theta_a(x) =
 \sum_{m=1}^\infty
\sqrt{g_a\over m} i\sin ({m\pi x \over L})
 ( b_{a m} - b_{a m}^\dagger )
+ \theta_a^{(0)}(x),
\end{equation}
\begin{equation}
\phi_a(x) =
\sum_{m=1}^\infty
\sqrt{1\over{g_a m}}\cos({m\pi x\over L})
 ( b_{a m} + b_{a m}^\dagger )
+ 2 \Phi_a
\end{equation}
where the zero-mode term $\theta_a^{(0)} \equiv {x\over{2L}} (\pi N_a +
\delta_a)$. The $b_{a m}$ satisfy
$[b_{a m},b_{a' m'}^\dagger] = \delta_{aa'}
\delta_{m m'}$ and
the operators $N_a$ and $\Phi_a$
satisfy $[N_a,\Phi_{a'}] = i\delta_{aa'}$.  Here we adopt
the notation $g_{\rho+} = g$ and $g_a = 1$ for the three neutral
sectors.
Substituting (5,6) into (2,4) the we may express the
Hamiltonian as ${\cal H} = \sum_a {\cal H}_a 
 - \mu N_{\rho+} -
\Delta_B N_{\sigma+} + \varepsilon_{\rm cap} N_{\rho-} $, where
we have included
a chemical potential $\mu$ controlled by external gates,
a Zeeman splitting $\Delta_B$ and
$\varepsilon_{\rm cap} = \pi v_F \delta_{\rho-}/4L$
\cite{phaseshift}.  Moreover,
\begin{equation}
{\cal H}_a =
{\varepsilon_a\over{8 g_a}} N_a^2
 +\sum_{m=1}^\infty m \varepsilon_a \  b_{a m}^\dagger b_{a m}
\end{equation}
where $\varepsilon_\rho = \pi \hbar v_\rho/L$ 
and $\varepsilon_a = \varepsilon_0 = \pi \hbar v_F/L$ for the
neutral sectors.

Consider the local tunneling density of states, $ A(x,\varepsilon) =
\sum_s |\langle s|\psi^\dagger(x)|0\rangle|^2 \delta(E_s - E_0 -
\varepsilon), $ which is proportional to $dI/dV$ measured in a
tunneling experiment.  This probes many body states $s$ in which one
electron has been added to the system at $x$.  The zero mode changes
$N_a$ by $\pm 1$ depending on the spin and band of the added
electron.  In addition, any number of collective modes may be excited.
Due to the structure of the bosonic excitation spectrum
and the fact that three of the four $\varepsilon_a$'s are equal, many
of these excited states will be degenerate.  $A(\varepsilon)$ thus consists
of a series of peaks,
\begin{equation}
A(\varepsilon) = \sum_{n_\rho n_0} C_{n_\rho n_0}
\delta(E_C + \varepsilon_0(n_0+{1\over 2}) +
\varepsilon_\rho n_\rho -\varepsilon )
\end{equation}
where,
$n_0$ and $n_\rho$
are non-negative integers and 
the Coulomb energy is given by
$E_C = (e^2/L)\ln R_s/R = \sum_a \pi \hbar v_a/8 L g_a - \varepsilon_0/2 $.
For simplicity we have set
$\mu= \Delta_B = \varepsilon_{\rm cap}=0$.
Since these terms couple only to the zero modes $N_a$, their
effect is to introduce a constant shift in energy of all of the peaks
for a given $N_a$.  Each of the peaks will thus in general 
be split into four by $\Delta_B$ and $\varepsilon_{\rm cap}$,  
and varying $\mu$ causes a constant shift in the energies of 
all of the peaks.

The amplitudes of the peaks in $A(x,\varepsilon)$
may be determined by computing the local Green's
function $G(x,t) = \langle\psi(x,0) \psi^\dagger(x,t)\rangle =
\int_0^\infty \! d\varepsilon\, 
A(x,\varepsilon) e^{i\varepsilon t}$.  Expressing $\psi(x)$ in terms of the
boson operators, this takes the form $G(t) = \prod_a {\cal G}_a(t)$,
where ${\cal G}_a(t) = \langle O_a(t)O_a^\dagger(0)\rangle$,
with $O_a = \exp[i(\phi_a\pm\theta_a)/2]$.
We then find
\begin{equation}
{\cal G}_a =
\left[{(\pi/L) e^{i\varepsilon_a t/2}  \over
{1 - e^{i\varepsilon_a t}}}\right]^{2 g_a^+}
\left[{{ 4\sin^2(\pi x/L) e^{i\varepsilon_a t}}
\over{(1-z e^{i\varepsilon_a t})
(1-z^* e^{i\varepsilon_a t})}}\right]^{g_a^-}.
\end{equation}
where $g^\pm_{a} = (g_a^{-1} \pm g_a)/16$ and $z = \exp 2\pi i x/L$.
By formally expanding $G(x,t)$ in powers of $e^{i\varepsilon_a t}$
it is then straightforward to extract the ratios,
\begin{equation}
C_{n_\rho n_0}/{C_{00}} = c_{3/4}^{n_0} \sum_{0\le i\le j \le n_\rho}
c_{2g^+}^{n_\rho-j} c_{g^-}^{j-i} c_{g^-}^i z^{j-2i}
\end{equation}
where $c_g^n = \Gamma(g+n)/\Gamma(g)\Gamma(1+n)$.

In Fig. 1 we plot the resulting density of states for tunneling into the end
and the middle of a tube for $\mu=B=\varepsilon_{\rm cap}=0$.
The tunneling spectrum is characterized by three energy scales.
As in the orthodox theory of the Coulomb blockade,
$E_C$ sets the minimum energy for adding an electron to
the tube.
The excited states fall into two categories:
The quantized spin/flavor excitations have energy
$\varepsilon_0 = \pi \hbar v_F/L$ - the unrenormalized
level spacing of the single electron states.  These
correspond to {\it neutral} collective excitations which are
unaffected by the long-range Coulomb interaction.   In addition, however, there
are charged ``quantized plasmon" excitations, which have an energy
$\varepsilon_\rho = \varepsilon_0/g = \sqrt{ \varepsilon_0
(8 E_C + \varepsilon_0)}$.  
This third energy scale is a signature of charge-spin separation - the hallmark
of a Luttinger liquid.
Since
$\varepsilon_\rho > \varepsilon_0$, the peaks in Fig. 1 fall into
distinct families with different plasmon excitations.  In particular,
the lowest family corresponds to tunneling into the ground state
of the charge sector.  The next family corresponds to exciting a
single quantum of the lowest energy plasmon: a dipole resonance.

The ratios of the peak heights contain detailed information about 
the interactions.  Within a given family,
$C_{n_\rho n_0}/C_{n_\rho 0} = c_{3/4}^{n_0}$, which is
$3/4$ for $n_0=1$ and approaches $n_0^{-1/4}/\Gamma(3/4)$ for $n_0\gtrsim 3$.  
The amplitude ratios between families depends on the tunneling location
$x$.  For the first plasmon excitation,
$C_{10}/C_{00} = 2(g_+ + g_- \cos(2\pi x/L))$.  Thus, as shown in
Fig. 1(b), the amplitude of the dipole resonance is suppressed when
tunneling into the middle of the tube.

\begin{figure}[htb]
  \setlength{\unitlength}{1.0in}
  \begin{picture}(4.0,5.0)(0.1,0)
    \put(0.3,2.7){\epsfxsize=3.0in\epsfbox{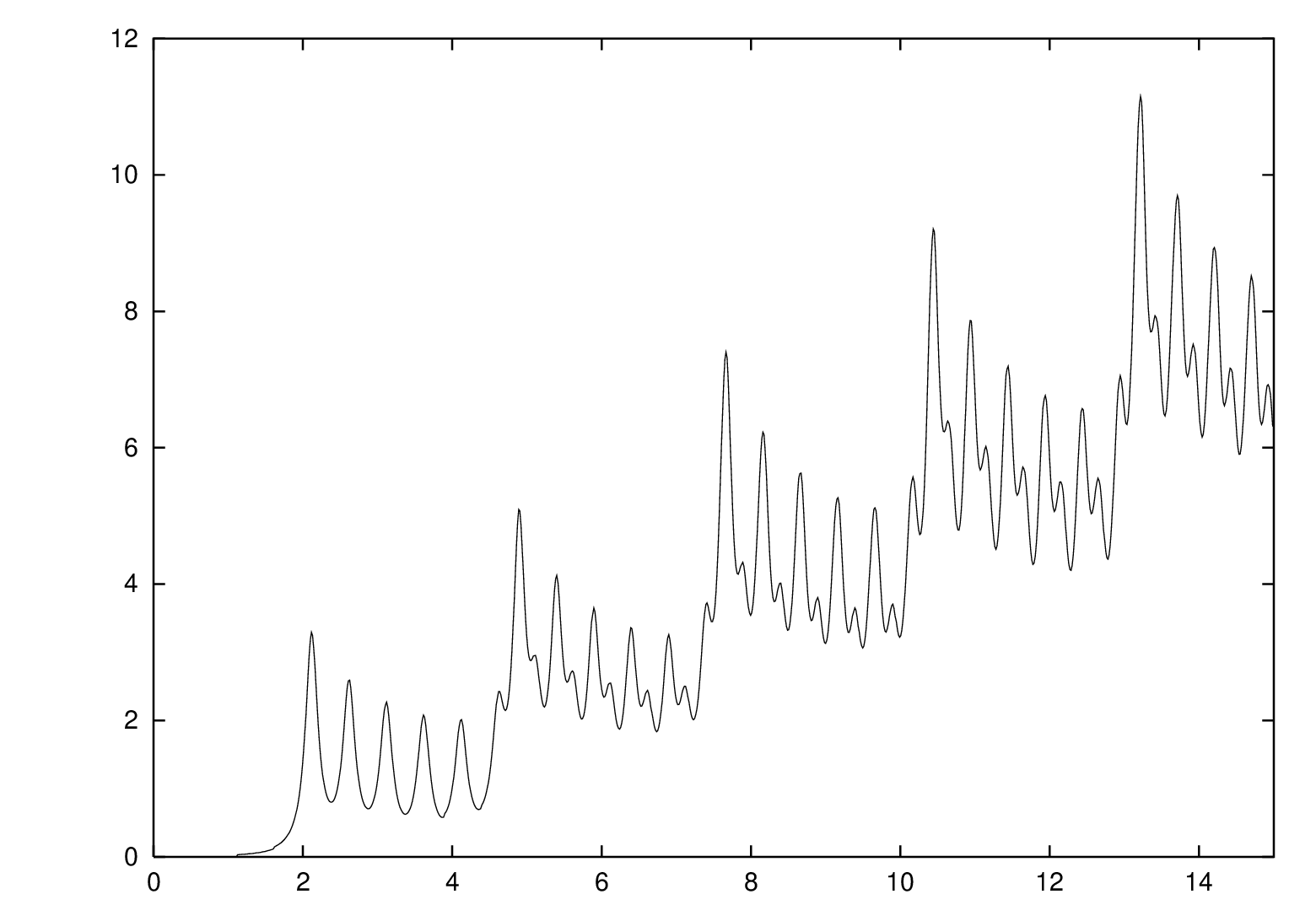}}
    \put(0.3,0.5){\epsfxsize=3.0in\epsfbox{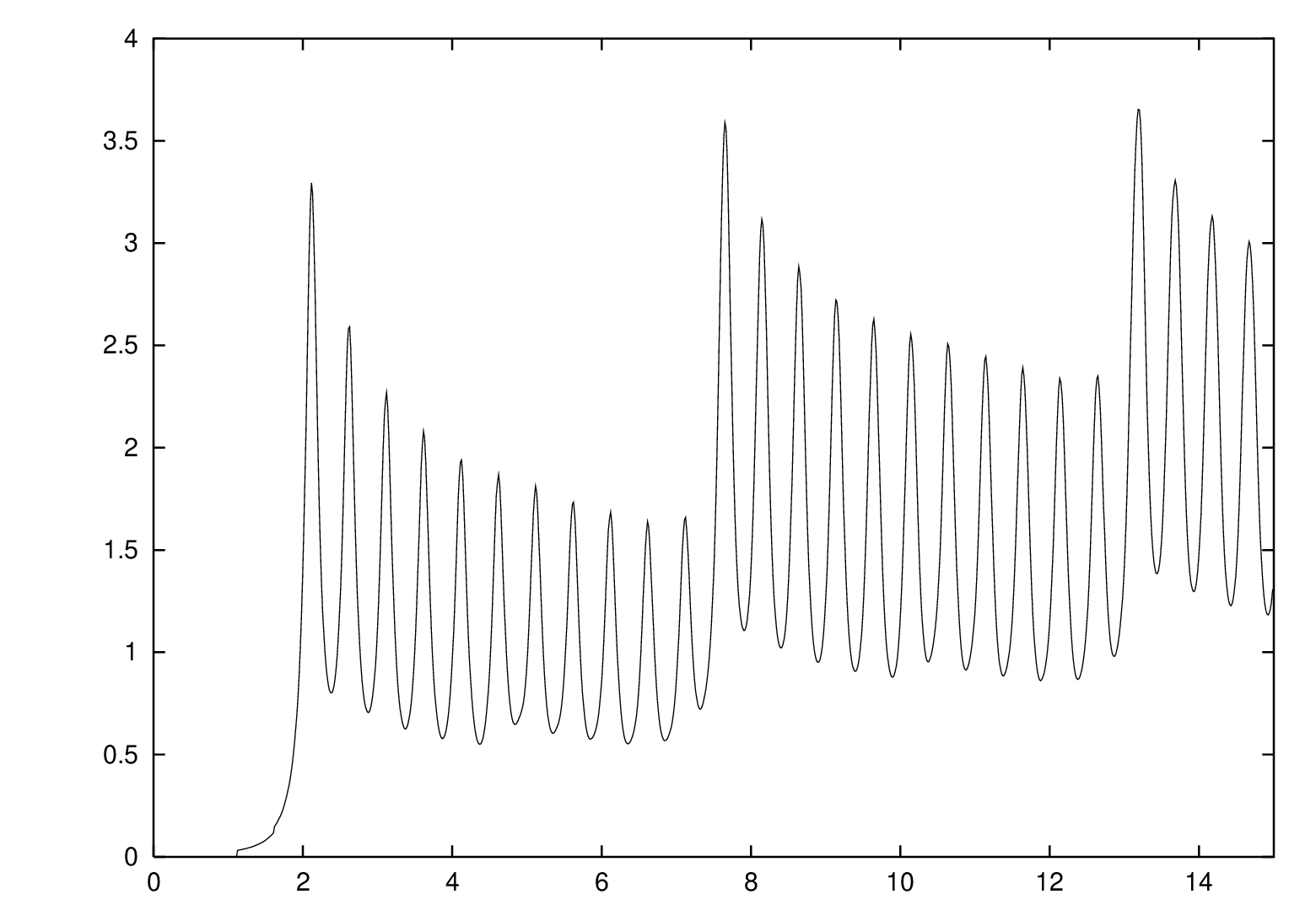}}
    \put(1.7,0.4){$\varepsilon$ (meV)}
    \put(1.7,2.6){$\varepsilon$ (meV)}
    \put(0.0,2.0){$A({L \over 2},\varepsilon)$}
    \put(0.0,4.0){$A(0,\varepsilon)$}
    \put(0.8,4.5){{\bf (a)}} 
    \put(0.8,2.3){{\bf (b)}}
  \end{picture}
  \caption{Local tunneling density of states {\bf (a)} at the end and
    {\bf (b)} in the center of the nanotube, shown for a nanotube of length
    $L=3 \mu m$ and $g=0.18$.  We have phenomenologically introduced a
    rounding to the spectral peaks, as would appear due to the Fermi
    distribution in the leads in a tunneling experiment.}
\end{figure}

It should be interesting in the future to explore both the low and
high temperature regimes experimentally.  We expect that in a fairly
clean experimental system, the power-law resistivity $\rho(T) \sim u^2
T^{2g-1} + n_0 T^{-(1-g)/2}$ is perhaps the most easily testable
prediction.  A potentially more rewarding experiment would be to
measure the tunnel conductance of an isolated nanotube with an STM, as
a function of bias, external gate potential, and position along the
tube, which we expect to be directly proportional to
$A(x,\varepsilon)$.
Clustering into families, as shown in the Figure, would give
direct and dramatic experimental evidence for the elusive charge-spin
separation of a Luttinger liquid.
%The predicted form requires, however 
%(1) negligible impurities; (2)
%sufficiently short tubes and/or doping such that $\Delta, \Delta_{bs}
%\ll \epsilon_0$; and (3) large-scale uniformity of the gate potential
%and other external perturbations.  
%Current measurements indeed observe Coulomb blockade behavior, but it is 
%more complex than predicted by
%the simple Luttinger model\cite{TunnelingExpts}.  
In addition to Coulomb blockade behavior, the current-voltage curves
observed by Tans et al. \cite{TunnelingExpts} 
display interesting structure, with signatures of a discrete
energy level spacing $\varepsilon_0 \approx .4$ meV
and an additional 2 meV step which may
be related to the lowest plasmon excitation.  
However, a detailed comparison with the simple Luttinger
model may be complicated, since the predicted form requires
(1) negligible impurities; (2)
sufficiently short tubes and/or doping such that $\Delta, \Delta_{bs}
\ll \epsilon_0$; and (3) large-scale uniformity of the gate potential
and other external perturbations.  
Nonetheless, a systematic study of the tunneling characteristics would 
be most useful at both low and high voltages and temperatures.
We encourage
mesoscopic experimentalists to rise to the challenge of tunneling
measurements in the Luttinger regime.

It is a pleasure to thank A.T. Johnson and E.J. Mele for helpful
discussions, and especially Cees Dekker for sharing experimental
results.  This work has been supported by the National Science
Foundation under grants No. PHY94-07194, DMR94-00142, DMR95-28578 and
DMR95-05425.


\begin{references}
\bibitem{phaseshift} 
These phase shifts depend upon the microscopic
structure of the cap, but are expected to be spin-independent.
\end{references}
\end{document}